# Analysis of M=1 Modes in The EAST Tokamak

Qing-Yun Hu(胡庆云)[1], Li-Qing Xu(徐立清)[2] and Dong-Jian Liu(刘东剑) [1**]

[1]College of Physics, Sichuan University, Chengdu 610065, China
[2]Institute of Plasma Physics, Chinese Academy of Sciences, Hefei 230031, China

An analysis of precession fishbone, diamagnetic fishbone and internal kink mode in Tokamak plasmas is presented via solving the fishbone dispersion relation. Applying the dispersion relation to a typical EAST discharge, excitation of precession fishbone due to Neutral Beam Injection is successfully explained. The real frequency and growth rate of diamagnetic fishbone and internal kink mode are calculated, and the relevance of the diamagnetic branch is also discussed for the possible equilibrium profile.

## 1. Introduction

Fishbone instabilities, excited by energetic particles (EPs) in magnetically confined plasmas, often result in loss of energetic particles, degrading the confinement of plasma and the heating efficiency. Since the first observation of fishbones in the Poloidal Divertor Experiment (PDX) tokamak in 1983 [1], two types of physics explanations have been proposed to explain the excitation of fishbone instabilities. The first theoretical model, given by *Chen et al.,* indicates that EPs could excite an energetic particle branch of (1, 1) mode when the (1, 1) internal kink mode is in its marginally stable regime. The (1, 1) EP branch, which has a real frequency comparable to the trapped particle toroidal precession frequency, is referred to as the precession fishbone [2,3]. The other model, taking diamagnetic frequencies into account, was suggested by *Coppi et al.* [4,5]. Since this mode has a real frequency related to ion diamagnetic frequency, it is referred to as diamagnetic fishbone. The discussion of these two branches was based on the marginal stability of the (1, 1) internal kink mode.

Fishbone oscillations have been observed in EAST neutral-beam-injection (NBI) experiments [6]. This branch of fishbone, with a frequency at the same order of precession frequency of trapped EPs, is the production of interaction between energetic particles and Internal Kink Modes. The coupling between EPs and Internal Kink modes not only drives fishbone modes but also degrades the confinement of energetic particles. Fishbone modes are often accompanied by a sawtooth crash in EAST, but the quantitative study of the modes in the EAST device is not sufficient. Because of the destructive effect of sawtooth on confinements, it is an urge to have an elaborate investigation on fishbones in EAST discharges.

In this paper, adopting real experiment data from an EAST discharge with NBI heating and using the combined dispersion relation from both fishbone models, we study the fishbone instabilities in the device. In part 2, we introduce the modified fishbone dispersion relation by considering the diamagnetic drift of and finite resistivity bulk plasma. Numerically solving the dispersion relation, we discuss the linear properties of the precession fishbone branch in section 3 and point out the possible relevance of the diamagnetic fishbone branch in the future EAST discharge.

## 2. Dispersion Relation

In this work, we treat the bulk plasmas as resistive MHD fluid, and the deeply trapped energetic ions produced by tangential injection of NBI as drift kinetic species under the assumption of large aspect ratio circular cross-section plasma equilibrium. With the generalized energy principle, the dispersion relation of fishbone modes can be written as

$$\delta\hat{I} + \delta\widehat{W}_c + \delta\widehat{W}_k = 0 \quad (1)$$

The first term $\delta\hat{I}$ is the normalized inertia of bulk plasma. $\delta\widehat{W}_c$ stands for the normalized ideal MHD potential energy variation of background plasma, and the nonadiabatic kinetic contributions from EPs are represented by $\delta\widehat{W}_k$.

In the case where finite ion gyro-radius and drift wave frequency effects are retained, $\delta\hat{I}$ reads [7]

$$\delta\hat{I} = -\frac{8i\Gamma\left[\left(\Lambda^{3/2}+5\right)/4\right]}{\Lambda^{\frac{9}{4}}\Gamma\left[\left(\Lambda^{3/2}-1\right)/4\right]}\frac{\left[\omega(\omega-\omega_{*i})\right]^{\frac{1}{2}}}{\omega_{A\theta}} \quad (2)$$

Where $\Lambda = \frac{-i\left[\omega(\omega-\omega_{*i})(\omega-\widehat{\omega}_{*e})\right]^{1/3}}{S_M{}^{-1/3}\omega_{A\theta}}$. $S_M$ denotes the magneto Reynolds number, $S_M = \tau_R/\tau_A$, $\tau_R = 4\pi r_s^2/\eta c^2$, $\tau_A = \omega_{A\theta}^{-1}$. $\omega_{A\theta}$ is the shear Alfvén frequency, $\omega_{A\theta} = \frac{B_\theta s}{r_s\sqrt{4\pi\rho}}$, $B_\theta, s, \rho$ are poloidal magnetic field, magnetic shear, mass density, respectively. $\omega_{*i}$ is the ion diamagnetic frequency. $\widehat{\omega}_{*e} = \omega_{*e} + 0.71\frac{c}{eBr_s}\left(\frac{dT_e}{dr}\right)_\eta$ with $\omega_{*e}$ the electron drift wave frequency. Note that this term is valid under the condition $Re\left(\Lambda^{3/2}\right) > -3$ and $Re\left[-iS_M{}^{-1}\omega_{A\theta}^{-1}\omega(\omega-\omega_{*i})/(\omega-\widehat{\omega}_{*e})\right]^{1/2} > 0$ [5,8].

For a parabolic safety factor q profile, $\delta\widehat{W}_c$ is calculated to be $3\pi\left(\frac{r_s}{R_0}\right)^2(1-q_0)\left(13/144-\beta_p^2\right)$ with $\beta_p = -\frac{2\mu_0}{B_\theta^2(r_s)}\int_0^1 dx\, x^2 dP/dx$ [9]. $R_0$ is the major radius. $q_0$ is the safety factor at magnetic axis. $x = r/r_s$.

For deeply trapped energetic ions with a slowing-down distribution $F_{h0} = n(r)\delta(\alpha-\alpha_0)E^{-3/2}$, the nonadiabatic kinetic contribution is expressed as [10–12]

$$\delta\widehat{W}_k = \beta_h\frac{1}{\varepsilon}\frac{\omega}{\omega_{dm}}ln\left(1-\frac{\omega_{dm}}{\omega}\right) \quad (3)$$

With $\beta_h$ denoting the spatial average of trapped particle $\beta$ within *q=1*. $\varepsilon$ is the inverse aspect ratio evaluated at $q = 1$ surface.

Now we can rewrite the dispersion relation as follows:

$$-\frac{8i\Gamma\left[\left(\Lambda^{3/2}+5\right)/4\right]}{\Lambda^{\frac{9}{4}}\Gamma\left[\left(\Lambda^{3/2}-1\right)/4\right]}\frac{\left[\omega(\omega-\omega_{*i})\right]^{\frac{1}{2}}}{\omega_{A\theta}} + \beta_h\frac{1}{\varepsilon}\frac{\omega}{\omega_{dm}}ln\left(1-\frac{\omega_{dm}}{\omega}\right) + \delta\widehat{W}_c = 0 \quad (4)$$

In the ideal MHD limit of bulk plasmas: $S_M \gg 1$, i.e. $\Lambda \gg 1$ (precession fishbone meets this condition well), equation (4) reduces to:

$$[\omega(\omega-\omega_{*i})]^{1/2} = i\omega_{A\theta}\left[-\delta\widehat{W}_c - \beta_h \frac{\omega}{\varepsilon\omega_{dm}} ln\left(1-\frac{\omega_{dm}}{\omega}\right)\right] \tag{5}$$

Near marginal stability, $\omega = \omega_{r,c} + i0^+$, and assuming $\omega_{*i} < \omega_{r,c} < \omega_{\mathrm{dm}}$, we have the real part and imaginary part of (5) separately [13] as:

$$[\omega_r(\omega_r-\omega_{*i})]^{1/2} = \pi\omega_{A\theta}\beta_{h,c}\frac{\omega_r}{\varepsilon\omega_{dm}} \tag{6}$$

$$-\delta\widehat{W}_c = \beta_{h,c}\frac{\omega_r}{\varepsilon\omega_{dm}} ln\left(\frac{\omega_{dm}}{\omega_r}-1\right) \tag{7}$$

Then, the relation between $\beta_{h,c}$ and $\delta\widehat{W}_c$ reads

$$-\delta\widehat{W}_c = \beta_{h,c}\frac{\omega_{*i}}{\varepsilon\omega_{dm}}\left[1-\left(\frac{\pi\beta_{h,c}\omega_{A\theta}}{\varepsilon\omega_{dm}}\right)^2\right]^{-1} ln\left\{\frac{\omega_{dm}}{\omega_{*i}}\left[1-\left(\frac{\pi\beta_{h,c}\omega_{A\theta}}{\varepsilon\omega_{dm}}\right)^2\right]-1\right\} \tag{8}$$

This relation divides the $(\beta_h, \delta\widehat{W}_c)$ plane into two regions in Figure 1, one is the stable region below the curve, and the unstable one is above it.

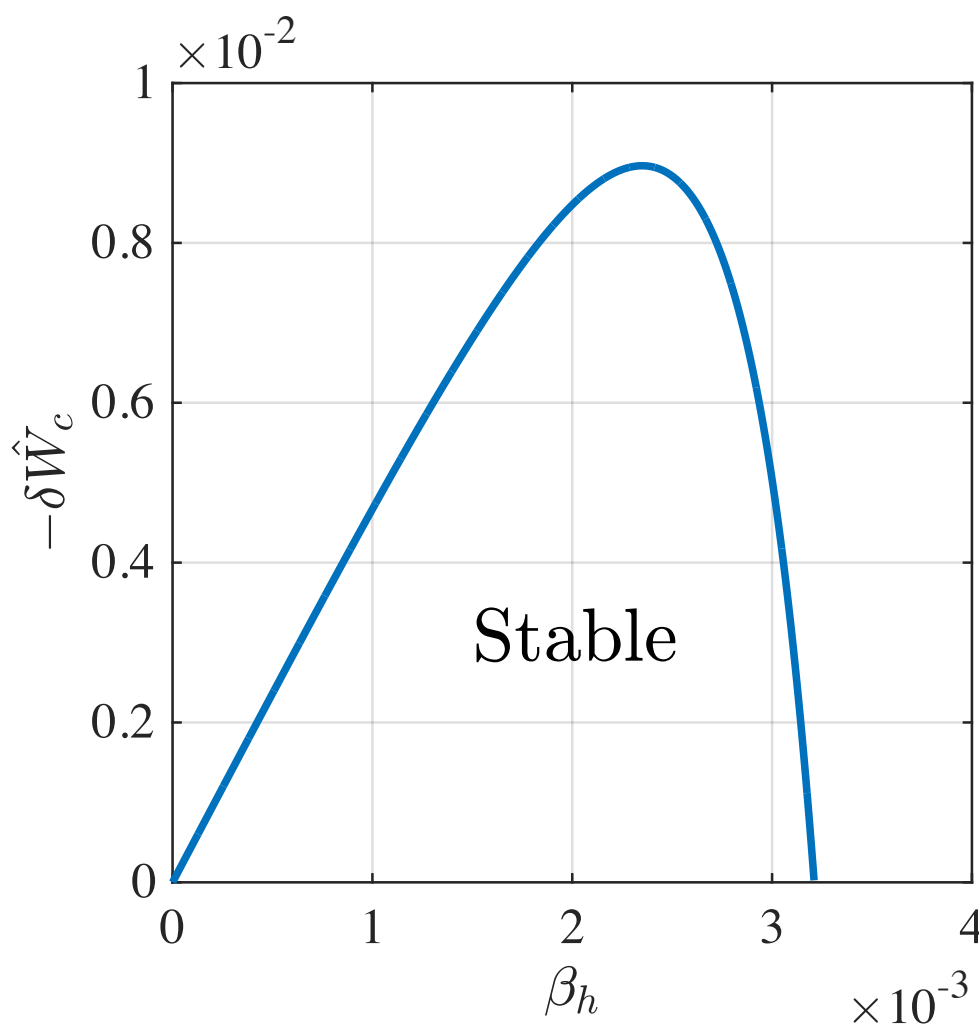

Figure 1. The stable domain for ideal fishbone modes in the $(\beta_h, \delta\widehat{W}_c)$ plane

## 3. Numerical Study

Using the above fishbone dispersion relation (4), let's analyze the typical fishbone signal in EAST discharge, and discuss the relevance of the two types of the fishbone branch, i.e. precession and diamagnetic fishbone in EAST.

### 3.1 Discharge Parameters

In EAST discharge: Shot #48605, the toroidal magnetic field $B_0 = 1.75T$; plasma current

$I_p = 400kA$; The neutral beam is injected perpendicularly with power $P_{NBI}$~$2.6MW$. The beta value of the neutral beam $\beta_{h,0}$~$0.005$, and the injection energy of the beam $E_i$~$64keV$. The major and minor radius are $R_0 = 1.86m$ and $a = 0.44m$, respectively.

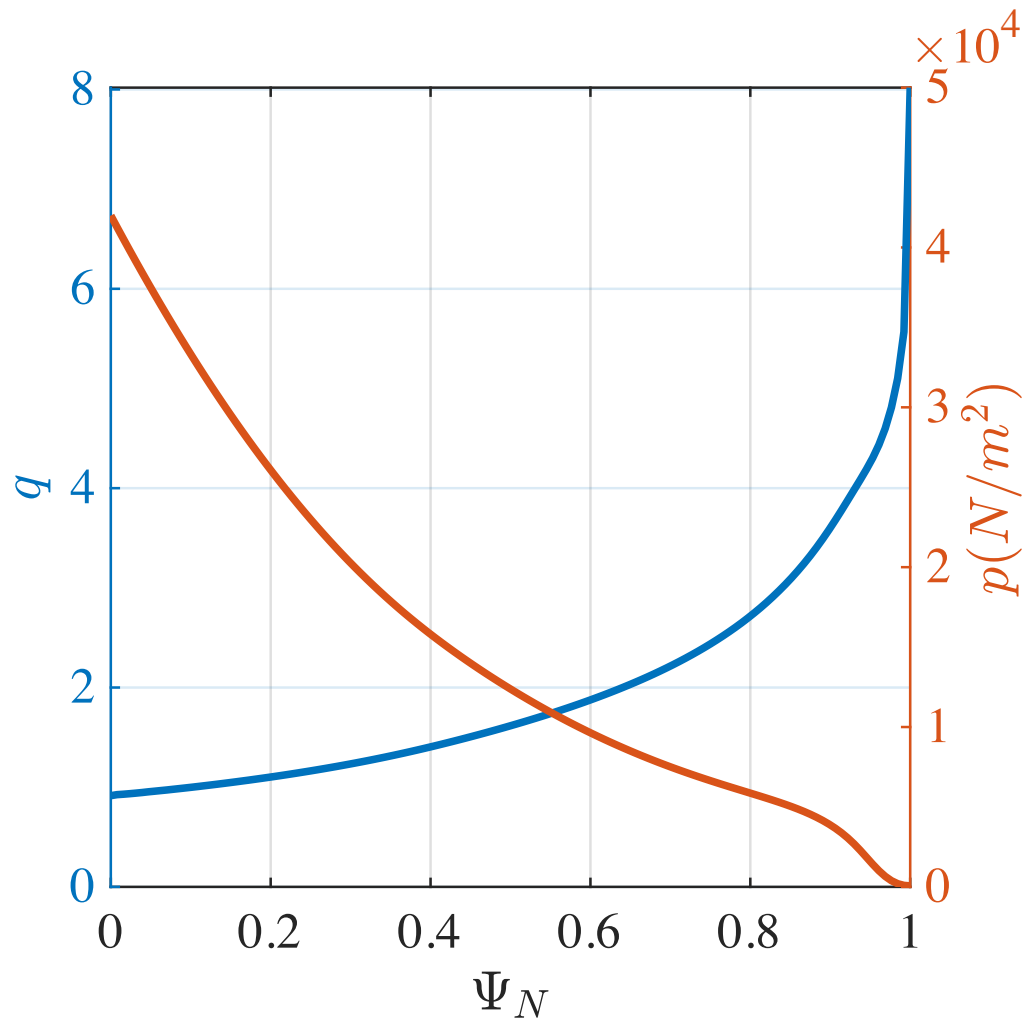

Figure 2. The profile of $q$ value and pressure $p$ vs. normalized poloidal magnetic flux $\Psi_N$ [6]

Figure 2 shows the equilibrium $q$ and pressure $p$ profile in magnetic flux coordinate obtained from the equilibrium data reconstructed with K-EFIT code. The central $q$ value $q_0 = 0.9150$, the location of $q$=1 surface $r_s = 0.107m$, the magnetic shear at $q$=1 surface $s = \frac{r_s}{q}\frac{dq}{dr} = 0.172$, the local poloidal magnetic field $B_\theta = \frac{r_s B_0}{R_0} = 0.101T$. The diagnostic data show that the electron density $n_{e0} = 5.28 \times 10^{19} m^{-3}$, bulk plasma beta $\beta_{total} = 0.0345$, plasma mass density $\rho = 1.77 \times 10^{-7} kg \cdot m^{-3}$. Shear Alfvén frequency is calculated to be $\omega_{A\theta} = 3.434 \times 10^5 rad/s$, bulk plasma poloidal beta $\beta_p = 1.1288$, bulk plasma potential energy perturbation $\delta\widehat{W}_c = -3.15 \times 10^{-3}$, the characteristic precession frequency of deeply trapped hot ions is $\omega_{dm} = 0.860 \times 10^5 rad/s$.

According to the definition of Spitzer resistivity $\eta = \frac{m_e \nu_{ei}}{e^2 n} = \frac{1}{(4\pi\varepsilon_0)^2} \frac{4\sqrt{2\pi} m_e^{1/2} Ze^2 \ln\Lambda_0}{3T_e^{3/2}}$, where $\Lambda_0 = \frac{\lambda_D}{b_0}$, $b_0 = \frac{q_e q_D}{4\pi\varepsilon_0 T_e}$, $\lambda_D = \sqrt{\frac{\varepsilon_0 T_e}{n_e e^2}}$, we have $\eta = 3.98 \times 10^{-8} \Omega \cdot m$ and the Magnetic Reynolds number $S_M = \frac{\mu_0 r_s^2 \omega_{A\theta}}{\eta} = 1.24 \times 10^5$.

### 3.2 Resistive Kink Mode

The dispersion relation i.e. equation (4) was solved numerically with an iterative solver in the complex eigenvalue plane, adopting parameters of an NBI discharge from EAST Tokamak. Given different initial guesses of the eigenvalue, the solver found different solutions. In this section, we focus on the pure MHD modes.

Near the ideal MHD marginal stability $\delta\widehat{W}_c = 0$, equation (4) yields root at $\omega = 0, \omega_{*i}$ or $\Lambda^{3/2} = 1, -3, -7, \cdots$ [14] Assuming $\Lambda^{3/2} \cong 1$, the dispersion relation reduces to [7,15,16]

$$-iS_M^{-1}\omega_{A\theta}^3 = \omega(\omega - \omega_{*i})(\omega - \widehat{\omega}_{*e}) \tag{15}$$

For the resistive kink mode, in the limit $|\omega| \gg \omega_{*i} = -\widehat{\omega}_{*e}$, the scaling of its growth rate is found to be $\gamma \propto S_M^{-1/3}$; if $|\omega| \ll \omega_{*i}$, the scaling becomes $\gamma \propto S_M^{-1}$. With the nonadiabatic kinetic contributions from hot ions $\delta\widehat{W}_k$ assumed to be 0, two roots of interest are discussed in this work: one root with $\omega_1 = (2.2 - 0.023i) \times 10^4 rad/s$ and the other with $\omega_2 = (-1.2 + 5.2i) \times 10^2 rad/s$.

Figure 3 compares the relation between growth rates of the two roots versus magnetic Reynolds number $S_M$. Under the condition $\omega_{*i} = 2.2 \times 10^4 rad/s$, the root $\omega_1$ and $\omega_2$ will be both rendered marginally stable as $S_M$ increases. For root $\omega_2$, the relation fits the scaling of $y = C_0 \times S_M^{-1}$, showing the destabilizing effect of resistivity on the mode and indicating this branch is the resistive kink mode. In addition, the scaling of (1, 1) resistive kink mode verified the validation of the numerical method we used in the following discussion.

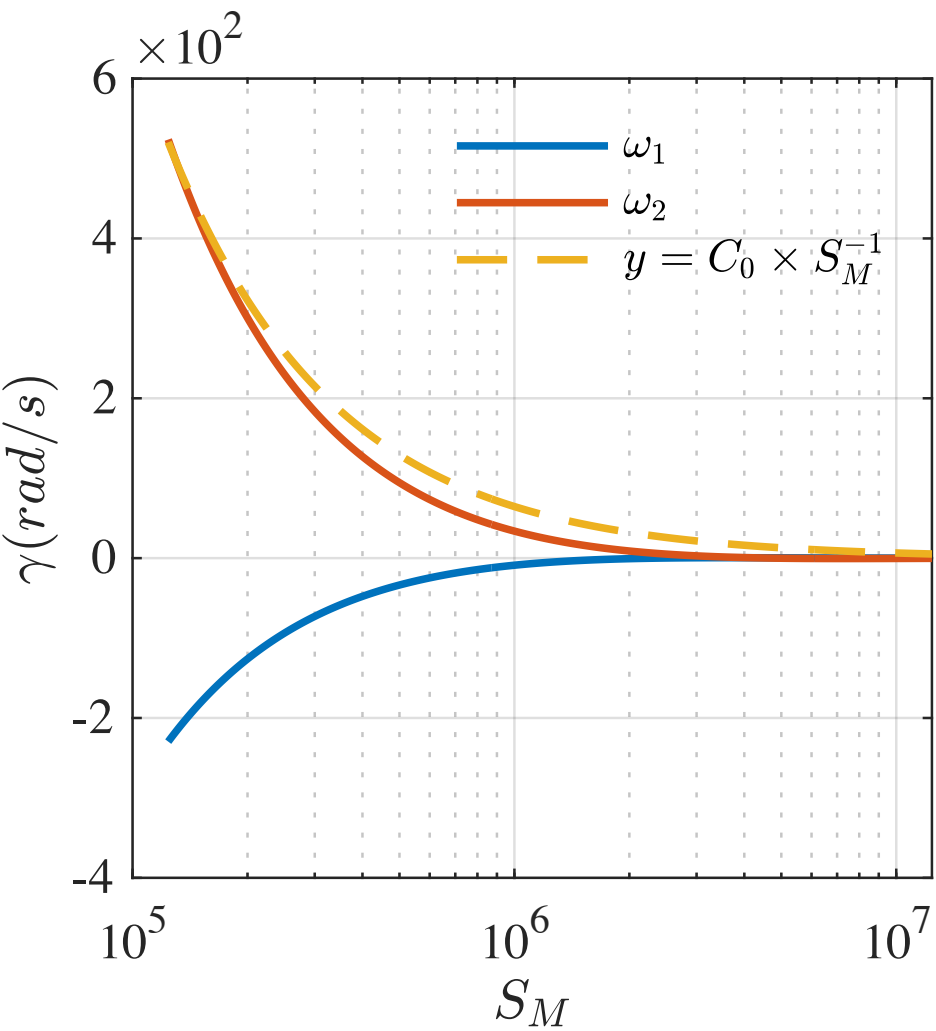

Figure 3. Growth rate of the two roots versus the magnetic Reynolds number $S_M$ and the fitting curve with function $y = C_0 \times S_M^{-1}$, where $C_0 = 6.5 \times 10^7$rad/s

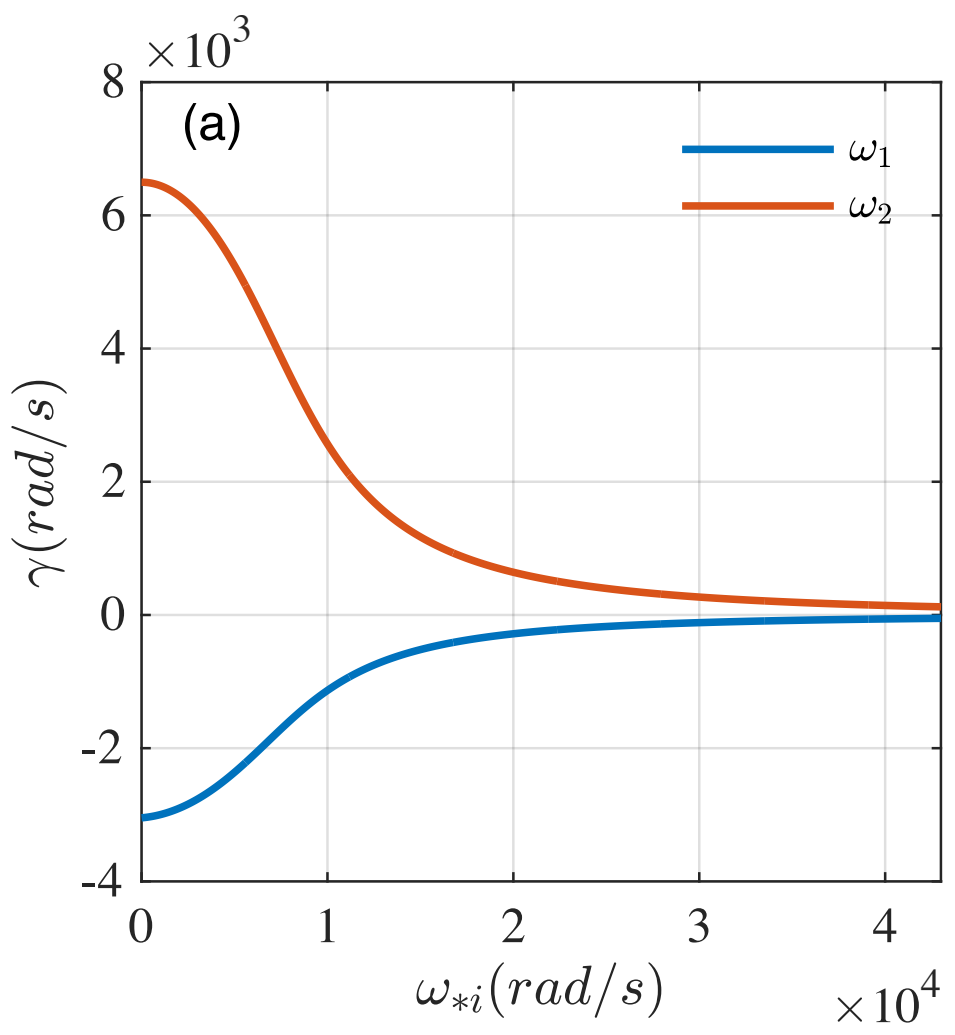

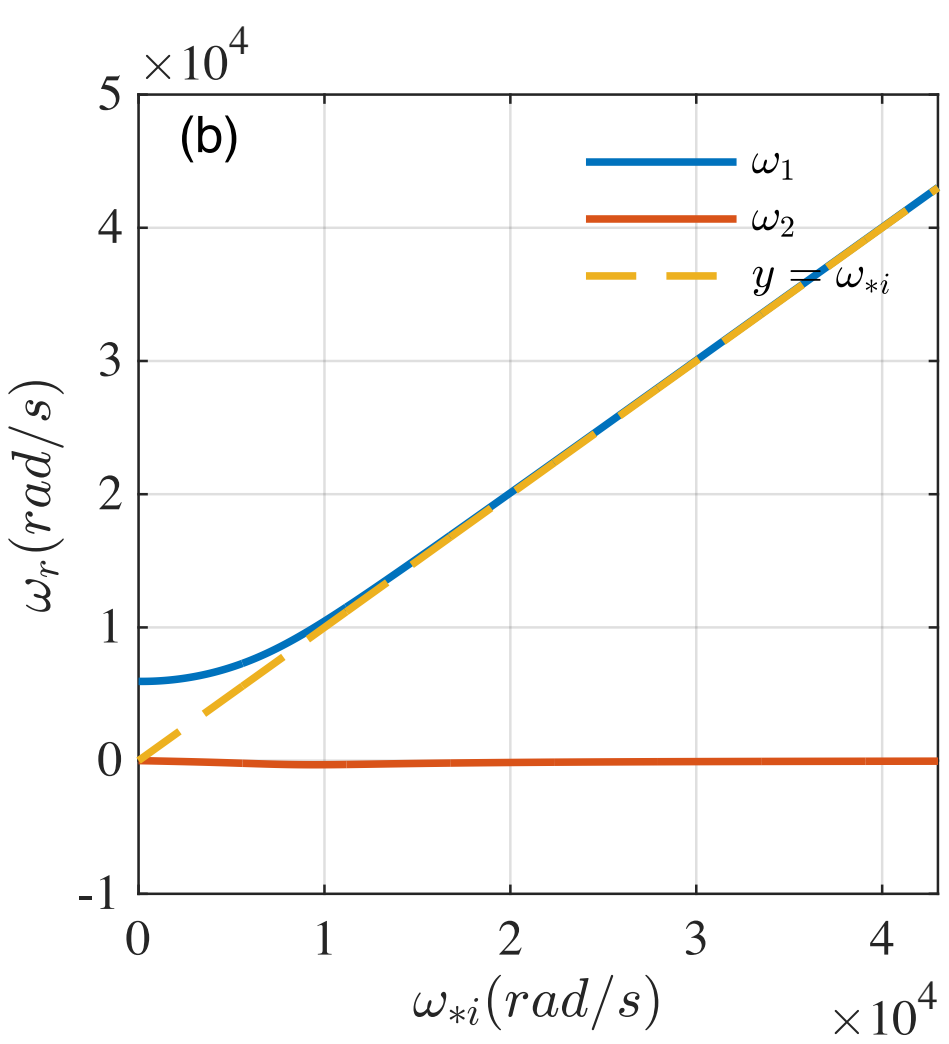

Figure 4. (a) Growth rate of the two roots versus $\omega_{*i}$; (b) Real frequency of the two roots versus $\omega_{*i}$ and the fitting curve with function $y = \omega_{*i}$

As to the diamagnetic drift effect on the two branches, one can conclude from Figure 4 that, given a specified q-profile and $\beta_p$, which means fixed free energy, the resistive kink mode will be stabilized as the increasing of diamagnetic frequency which stored free energy of system into the diamagnetic term.

Turning on the $\delta\widehat{W}_k$ term and assuming $\omega_{*i} = -\widehat{\omega}_{*e} = 2.2 \times 10^4 rad/s$, the root $\omega_2$, which we define as resistive kink mode is studied in the presence of perpendicular NBI injection. Figure 5 shows that, the “dissipation” from hot ions has a stabilizing effect on the mode. The orderings of its growth rate are also shown in equation (14) [4,12]. Also, the drop of free energy in bulk plasma or increase of diamagnetic frequency stabilizes the mode.

$$\gamma_k \cong \frac{5}{2} \frac{\omega_{A\theta}^3}{S_M |\omega_{*i} \widehat{\omega}_{*e}|} + 2\pi \frac{\beta_{\mathrm{h}}}{\varepsilon} \frac{\delta \widehat{W}_c^3 \omega_{A\theta}^4}{\omega_{dm} \omega_{*i}} \tag{14}$$

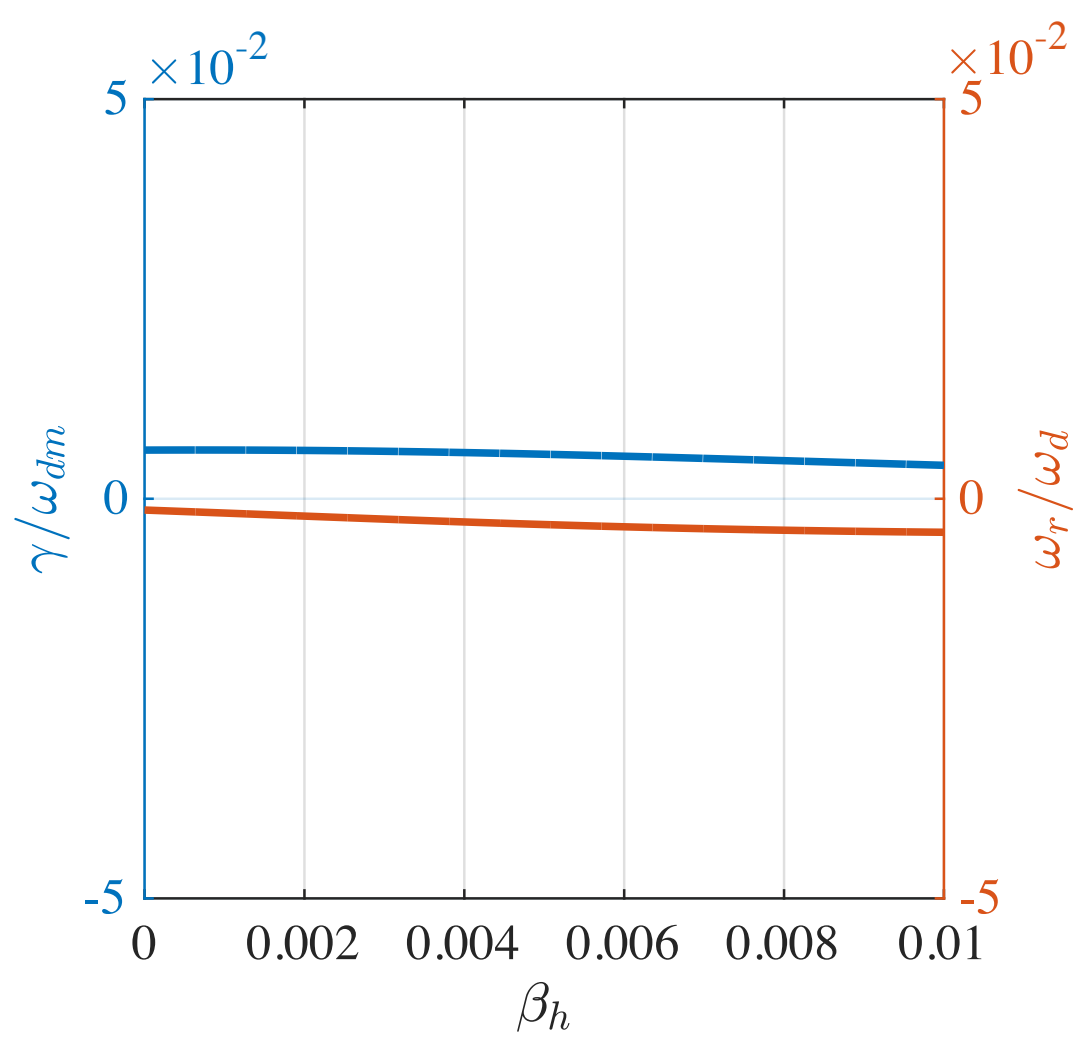

Figure 5. The real frequency and growth rate (both normalized by $\omega_{dm}$) of resistive kink mode versus $\beta_h$ with $\omega_{*i} = -\widehat{\omega}_{*e} = 2.2 \times 10^4 rad/s$ under the real discharge parameters

### 3.3 Precession Fishbone

In this section, we focus on the precession fishbone in the same discharge. First, we investigate the influence of magnetic Reynolds number $S_M$ on the precession fishbone. Then we analyze this branch in ideal limit near marginal stability, giving its real frequency $\omega_r$, critical beta $\beta_{h,c}$ and growth rate $\gamma_p$ approximately. Finally, numerical results of various equilibrium parameter effects on precession fishbone are discussed in detail.

Relation between the threshold beta value of precession fishbone $\beta_{h,c}$ and $S_M$ is shown in Figure 6 by changing the magnetic Reynolds number $S_M$ in range $10^2 - 10^8$,. The critical $\beta_{h,c}$ of the precession fishbone branch increases sharply near $S_M{\sim}10^3$ which implies it is difficult to excite the branch during the ohmic discharge. As the increasing of the plasma temperature, the resistivity of the background plasmas becomes smaller, bulk plasma can be treated as ideal MHD background, and the critical $\beta_{h,c}$ tends to flatten after $S_M{\sim}10^5$, which can also be seen from Figure 6. Similarly, the real frequency near marginal stability is relatively large during ohmic discharge and then flatten after $S_M$ exceeding $10^5$.

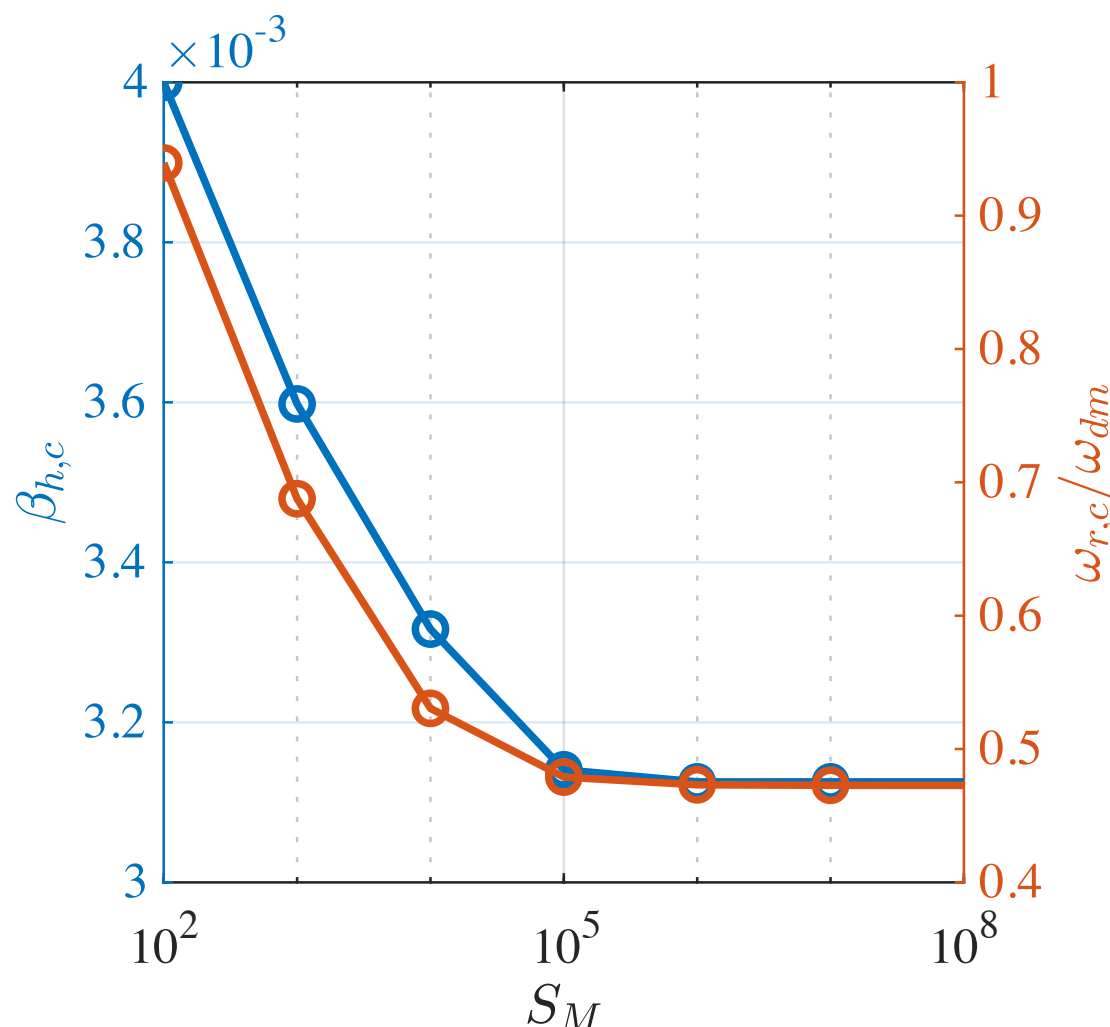

Figure 6. The threshold beta value of precession fishbone $\beta_{h,c}$ and real frequency of precession fishbone near marginal stability versus the magnetic Reynolds number $S_M$

From equations (6) and (7), we find an asymptotic solution of $\omega_{r,c}$ for this branch

$$\omega_{r,c} \cong \frac{\omega_{dm}}{2}\left(1 - \left|\frac{\delta\widehat{W}_c \pi \omega_{A\theta}}{\omega_{dm}}\right| \Big/ \sqrt{1 - \frac{2\omega_{*i}}{\omega_{dm}}}\right) \tag{9}$$

And get the critical value of $\beta_h$ for small $\omega_{*i}$ from equation (6)

$$\beta_{h,c} \cong \frac{\varepsilon \omega_{dm}}{\pi \omega_{A\theta}}\left[1 - \frac{\omega_{*i}}{\omega_{r,c}}\right]^{1/2} \tag{10}$$

Hence, the mode will be easier to be excited as finite ion diamagnetic frequency increases. Under the condition of $\delta\widehat{W}_c \pi \omega_{A\theta}/\omega_{dm} \leq O(1)$, $\beta_h \cong \beta_{h,c}$, the growth rate of precession fishbone is approximated as

$$\gamma_p \cong \frac{\pi^2 \omega_{A\theta}}{4\varepsilon}\left(\beta_h - \beta_{h,c}\right) \tag{11}$$

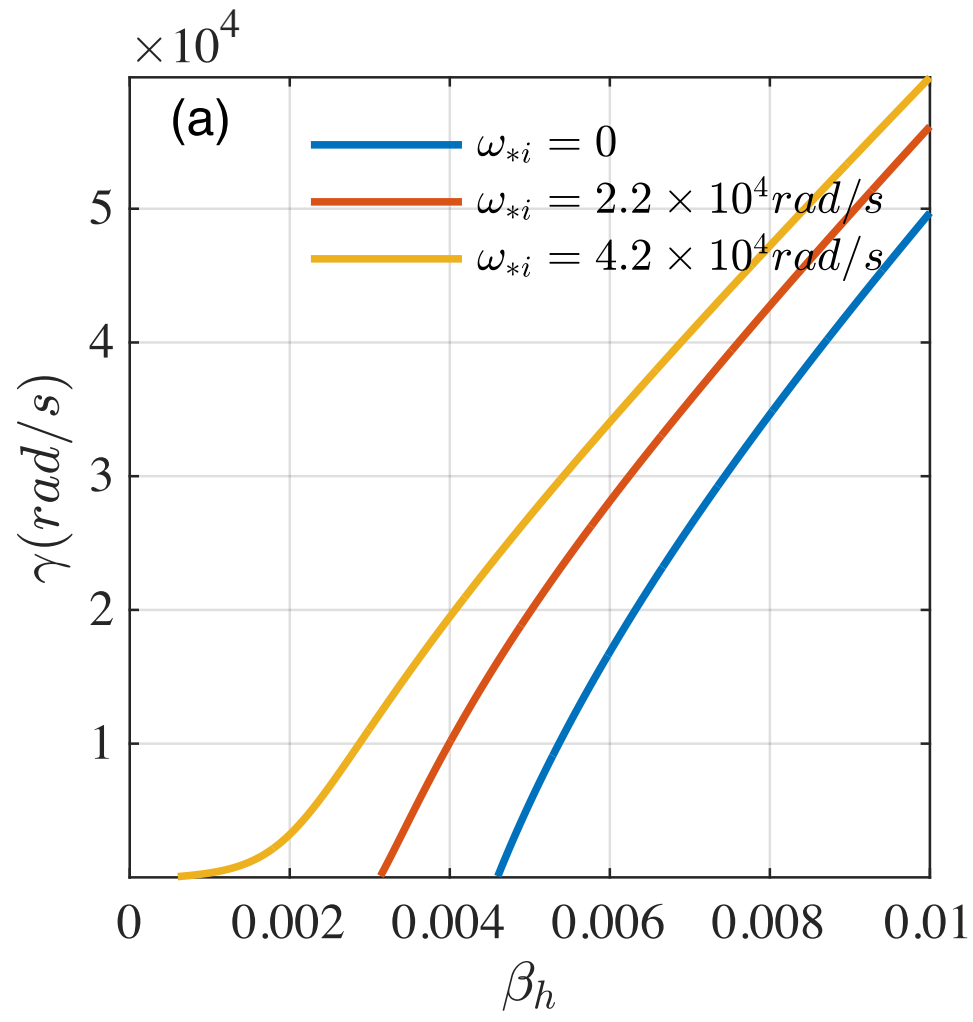

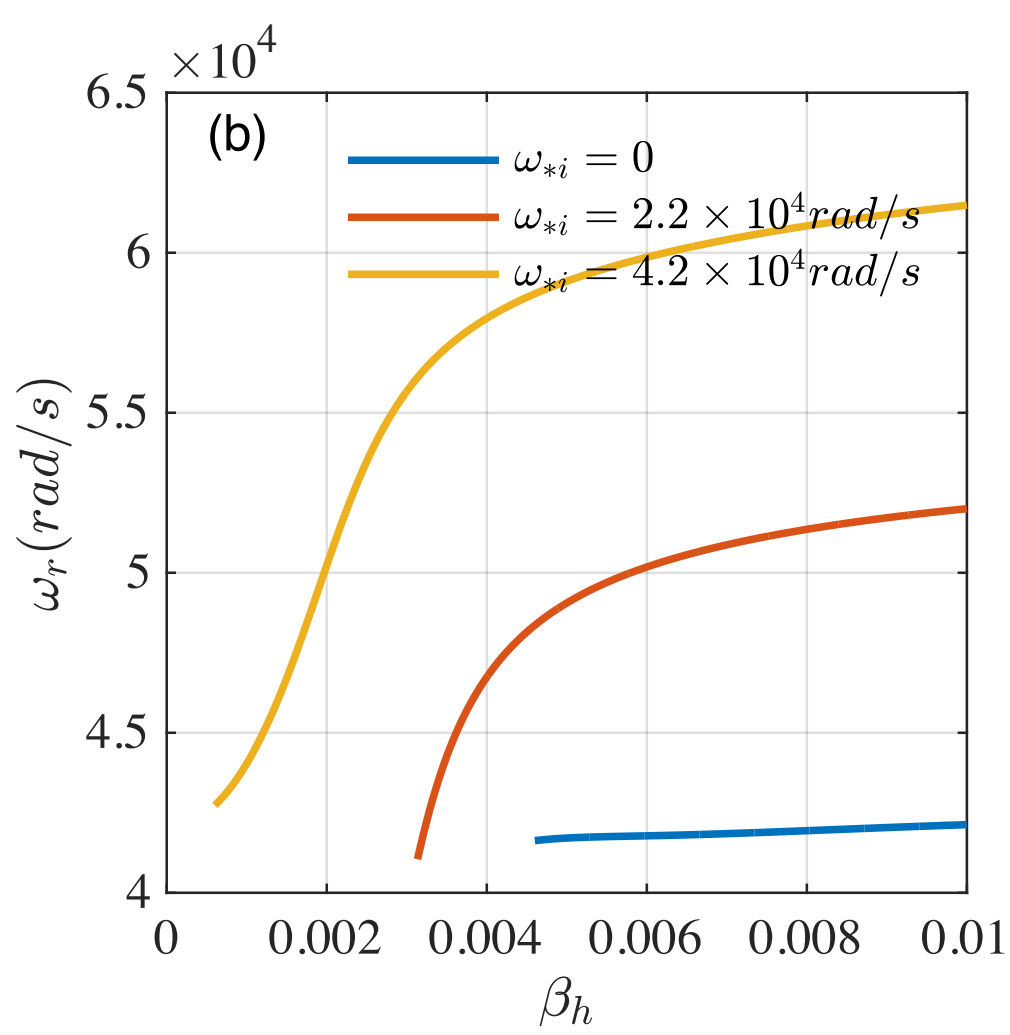

Figure 7. (a) Growth rate and (b) Real frequency of Precession Fishbone versus $\beta_h$ with different ion diamagnetic frequencies

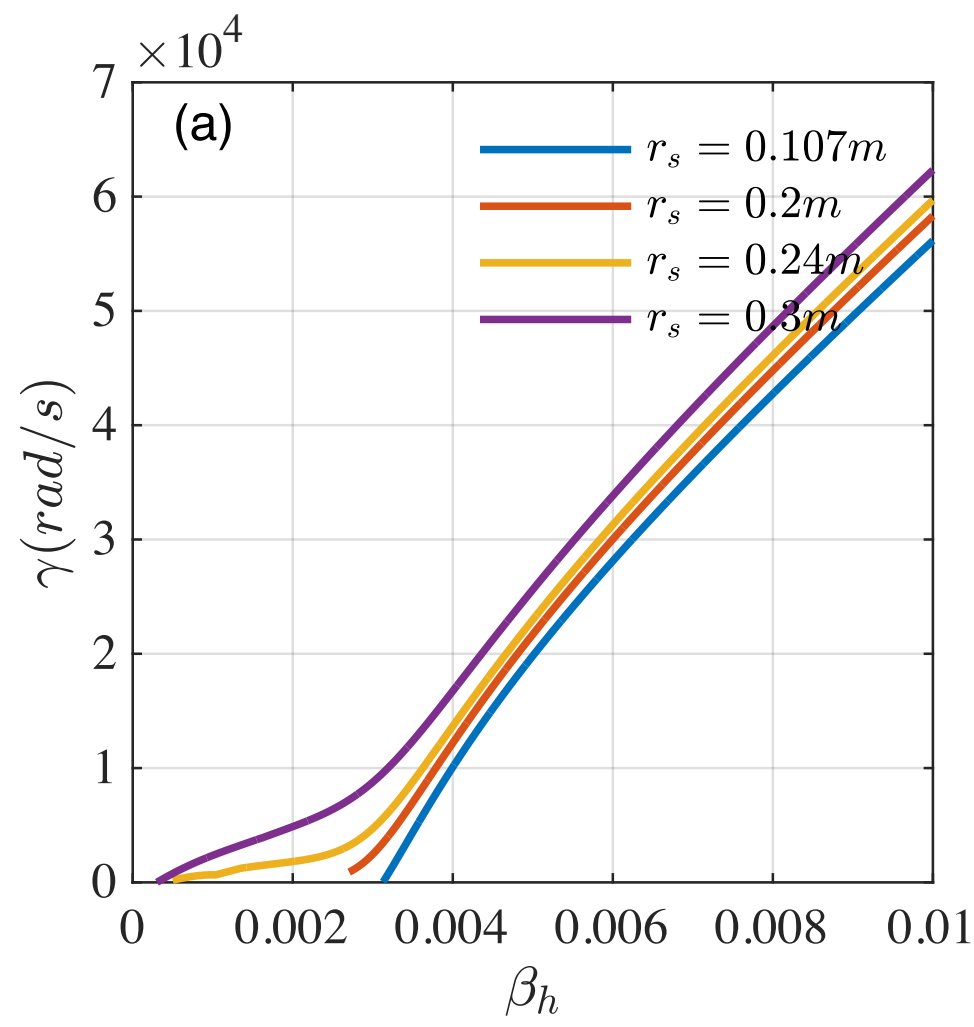

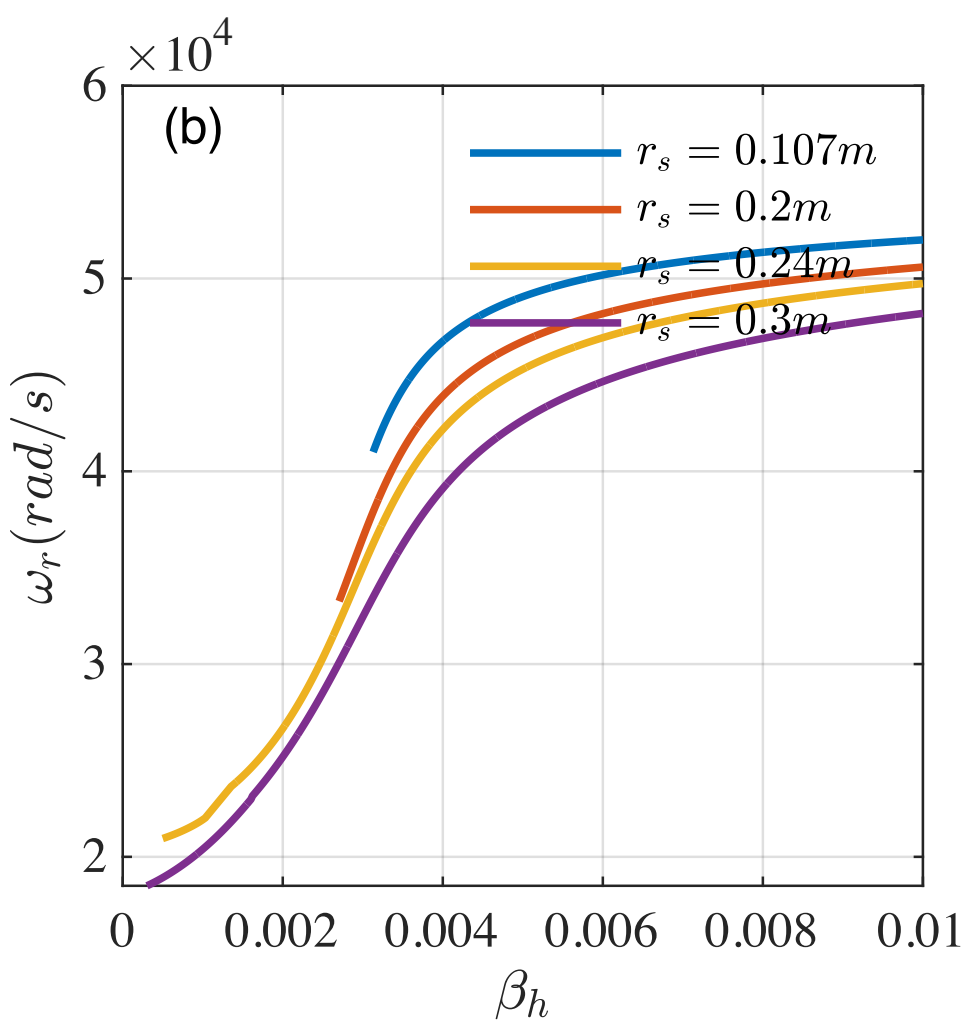

Figure 8 (a) Growth rate and (b) Real frequency of Precession Fishbone versus $\beta_h$ with different $r_s$ in $\delta\widehat{W}_c$ term

Figure 7 (a) and (b) represent the variation of growth rate and real frequency of the mode versus $\beta_h$ with different $\omega_{*i}$. Assuming $\omega_{*i} = -\widehat{\omega}_{*e} = 0$, the mode is excited when $\beta_h = 0.0046$, i.e. $\beta_h/\beta_{total} = 0.13$. When $\beta_h/\beta_{total} = 0.25$, $\gamma_p$ is about $0.4620\omega_{dm} = 3.97 \times 10^4 rad/s$, $\omega_r$ is around $0.4894\omega_{dm} = 4.21 \times 10^4 rad/s$. The results match well with simulation results [17]. With $\omega_{*i} = 2.2 \times 10^4 rad/s$, $\widehat{\omega}_{*e} = -\omega_{*i}$, $\omega_{dm} = 0.860 \times 10^5 rad/s$, the mode is excited when $\beta_h = 0.0031$, $\omega_r$ is around $(0.47 \sim 0.6)\omega_{dm}$.

As described by Equations (9) and (10), critical $\beta_{h,c}$ will be modified by finite $\omega_{*i}$, as $\omega_{*i}$ increases, $\beta_{h,c}$ keeps on decreasing. For $\omega_{*i} \ll \omega_{dm}/2$, the real frequency also decreases as $\omega_{*i}$ increases. In this regime, diamagnetic rotation plays a role in storing free energy. As $\omega_{*i}$ increases, the resonance between diamagnetic rotation and the mode makes the excitation easier. Note that under the condition $\omega_{*i} \approx \omega_{dm}/2$, it is shown that the precession fishbone is marginally stable with a real frequency $\omega_r \approx \omega_{*i} \approx \omega_{dm}/2$. When $\omega_{*i} > \omega_{dm}/2$, this branch vanishes. Resonance from diamagnetic rotation becomes dominant.

In Figure 8, we vary the $|\delta\widehat{W}_c|$ by changing the mode rational surface position $r_s$. The growth of $r_s$ is equivalent to the increase of $|\delta\widehat{W}_c|$, as indicated by equations (9) and (10), leads to the decrease of $\beta_{h,c}$ and $\omega_{r,c}$. The drop of free energy makes the mode easier to be resonantly destabilized.

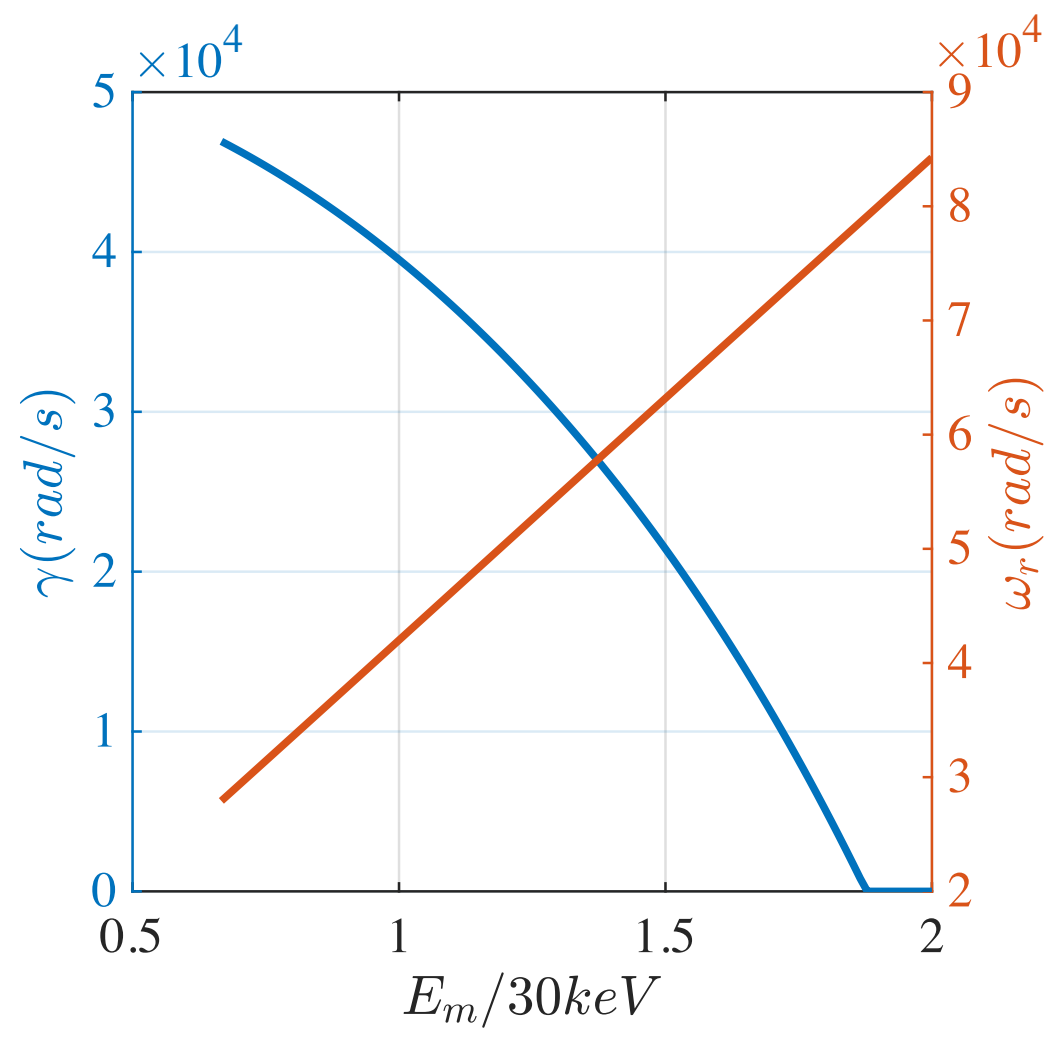

Figure 9. The real frequency (solid line) and growth rate (dotted line, both normalized by $\omega_{dm}$) of precession fishbone with $\beta_h/\beta_{total} = 0.25$ versus $E_m$ (normalized by 30keV)

Figure 9 shows the scaling of the real frequency and the growth rate of the mode with the characteristic energy of deeply trapped ions. We find that for a certain $\beta_h$, growth rate decreases as precession frequency increase, while real frequency increase on the contrary. The increase of $\omega_{dm}$ raises the mode frequency to meet the resonant condition as well as the critical $\beta_h$, which is indicated by equations (9) and (10). The free energy needed for mode growth becomes larger. Hence, for a fixed value of $\beta_h$, the growth rate will drop as the increase of $\omega_{dm}$.

### 3.4 Diamagnetic Fishbone

In this section, diamagnetic fishbone in the same discharge is investigated. This branch of fishbone is one of the two (1, 1) modes which need "viscosity" from EPs to develop [5]. We present the analytical expression of its growth rate and find this solution in numerical calculation.

Considering the ordering of $\beta_h$, the real part of $\delta\widehat{W}_k$ can be neglected. We have $\delta\widehat{W}_k = -i\pi\beta_h\frac{\omega}{\varepsilon\omega_{dm}}$. At the limit of $\omega \approx \omega_{*i}$, we have $\Lambda^{3/2} > 1$, which reads $(\omega_{A\theta}/S_M\omega_{*i})^{1/2} < Im(\delta\widehat{W}_k) + \delta\widehat{W}_c < \omega_{*i}/\omega_{A\theta}$. Assuming $|Im(\delta\widehat{W}_k)| < |\delta\widehat{W}_c|$, the dispersion relation gives [12]

$$\gamma_d \cong -2\pi\frac{\beta_h}{\varepsilon}\frac{\omega_{A\theta}^2}{\omega_{dm}}\delta\widehat{W}_c - \frac{5}{2}\frac{\omega_{A\theta}^3}{S_M\omega_{*i}(\omega_{*i} - \widehat{\omega}_{*e})} \tag{12}$$

Equation (12) indicates that "dissipation" from hot ions is needed for the mode to develop. Furthermore, for $\omega_{*i} < \omega_{dm}$, the increase of $\omega_{dm}$ have a stabilizing effect on the mode, because it makes resonance between hot particles and the mode more difficult.

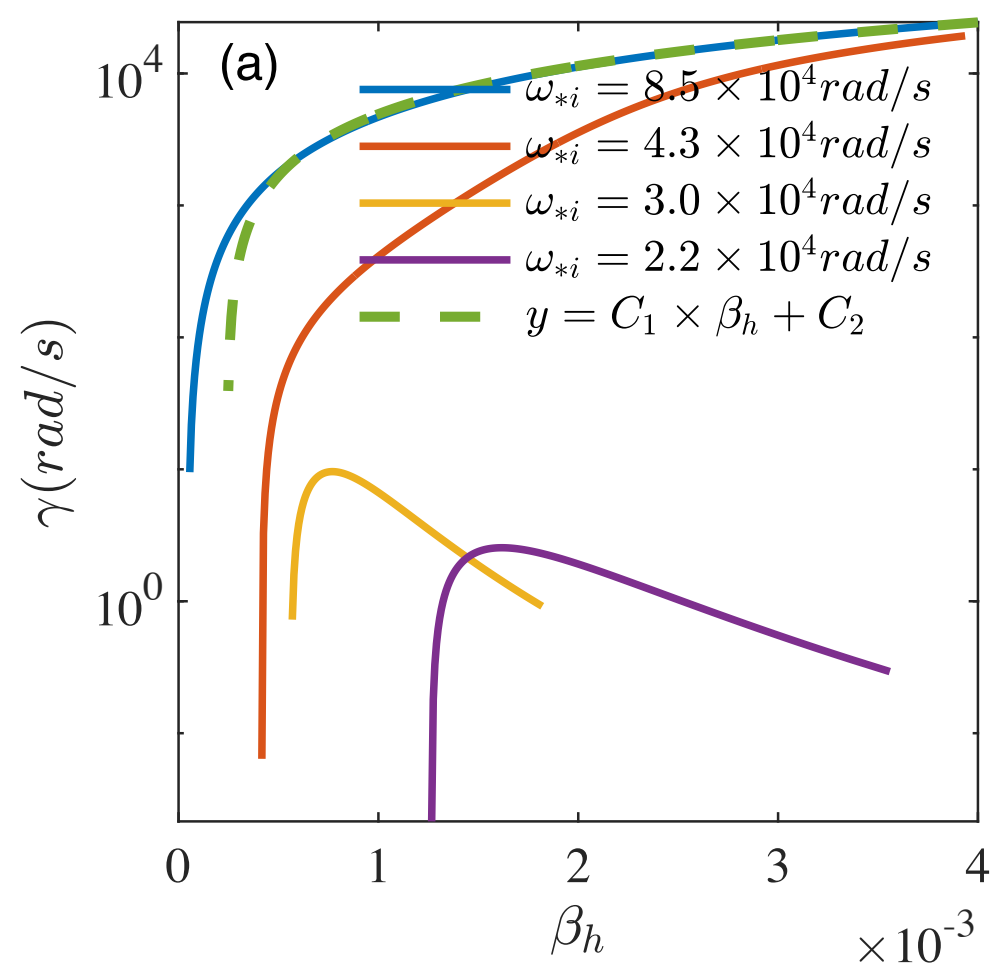

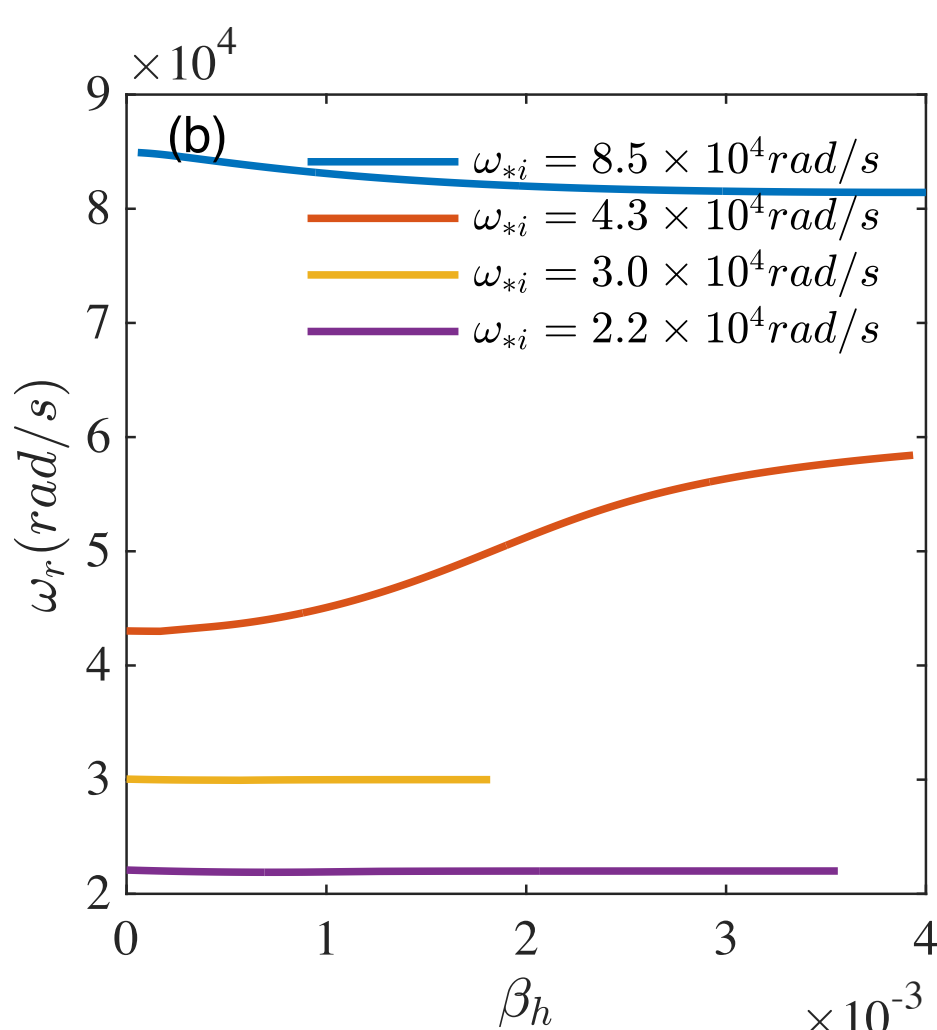

Figure 10. (a) Growth rate and (b) Real frequency of Diamagnetic Fishbone versus $\beta_h$ with different ion diamagnetic frequencies; $C_1 = 6.49 \times 10^6 rad/s$, $C_2 = -1.57 \times 10^3 rad/s$.

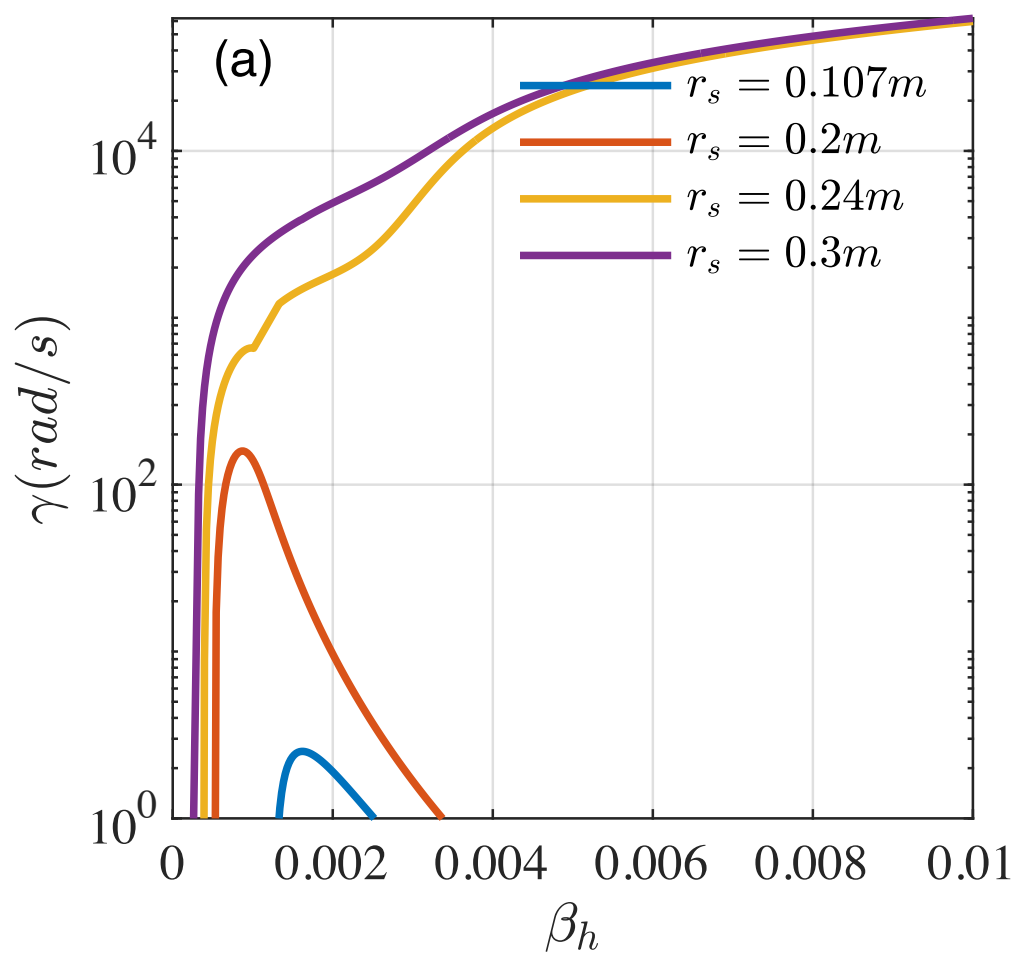

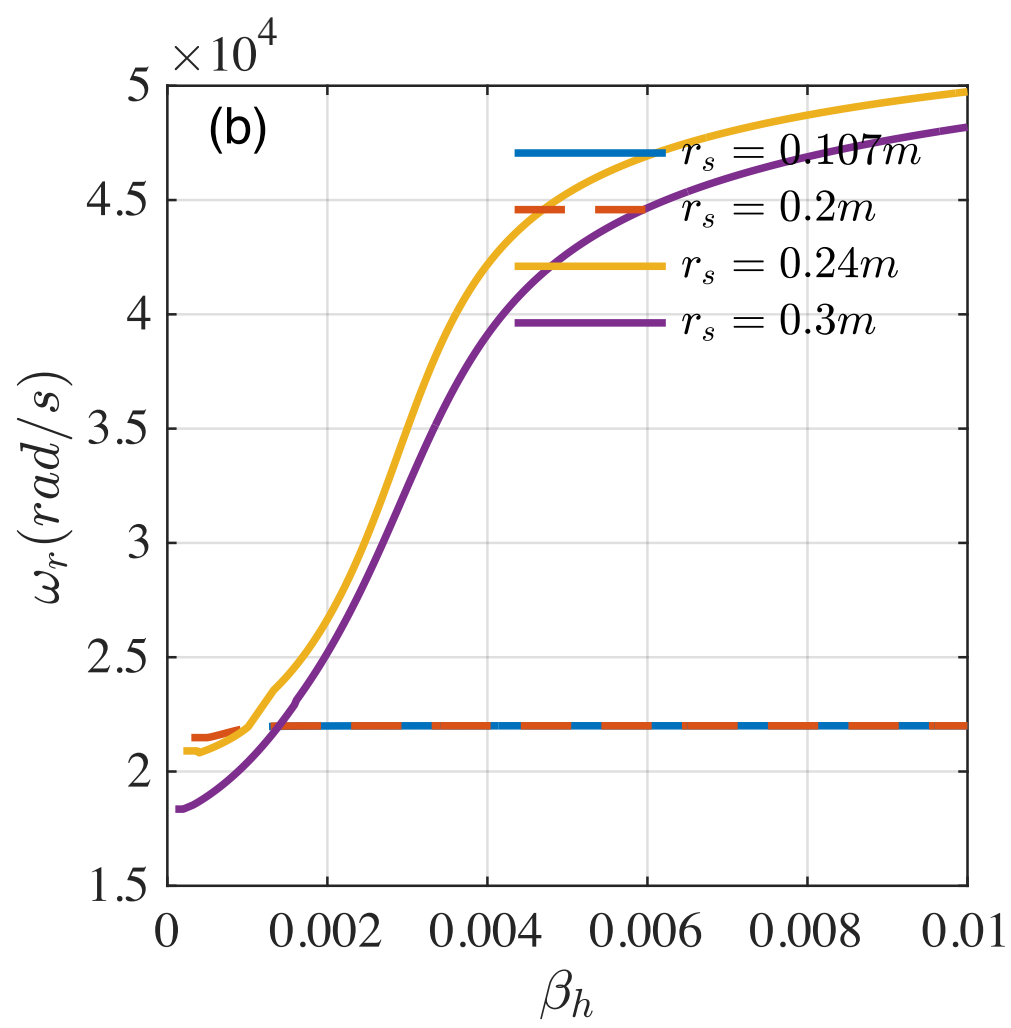

Figure 11 (a) Growth rate and (b) Real frequency of Diamagnetic Fishbone versus $\beta_h$ with different $r_s$ in $\delta\widehat{W}_c$ term

In Figure 10 and Figure 11, it can be seen that the growth rate of this root shows different behaviors as $\omega_{*i}$ and $r_s$ change. When $\omega_{*i}$ and $r_s$ is large enough, which indicates the pressure gradient of bulk plasmas is large while q-profile is flat, the growth rate depends positive-linearly on $\beta_h$ (as described in Equation 12). In this case, the inertia of ideal MHD fluid of the bulk plasmas meets its approximation $\Lambda^{3/2} \gg 1$, and the reduced form of Gamma function mentioned above is reasonable.

However, if $\omega_{*i}$ is not sufficiently large, the mode will be suppressed as $\beta_h$ increases and the value of $\Lambda^{3/2}$ will be smaller than the unit. The reduction in the ideal limit is not proper to be used.

## 4. Discussion

To have a complete understanding of the sawtooth physics picture in the EAST discharge, let the beta value of hot ions $\beta_h$ varies from 0 to 0.01, three roots of interest are shown in Figure 12. The first root, with $\omega_r \approx \omega_{dm}/2$ is the Precession Fishbone (PF) mode. It appears when $\beta_h$ exceeds the critical value $\beta_{h,c} = 0.0031$, which is lower than the experimental NBI beta value $\beta_{h,0} = 0.005$ of the experiment. The second root, with $\omega_r \approx \omega_{*i}$ is the Diamagnetic Fishbone (DF) mode. As $\beta_h$ increases, it keeps marginal stable with a growth rate much lower than the other two (about 4 magnitudes) in the regime $\beta_{h,c}\sim\beta_{h,0}$. The third root, with $\omega_r \approx 0$ is the Resistive Kink (RK) mode. The growth rates and the real frequency of the roots are shown in Figure 13. In the regime around $\beta_h = \beta_{h,0}$, the root PF has a growth rate that is larger than the other two. Furthermore, the diagnostic data of the discharge from Ref. [6] shows that the oscillation frequency is in the range $(1.75 - 5.75) \pm 1.25kHz$. From the SX signal of the oscillation, the maximum growth rate is calculated to be around $2.9 \times 10^3 rad/s$. Frequency and growth rate obtained from our calculation are in the range $(6.5 - 7.8)kHz$ and $(0 - 19.8) \times 10^3/s$, respectively. Our calculation results are basically consistent with experimental data. Thus, we can attribute the oscillations in this discharge to the Precession Fishbone mode.

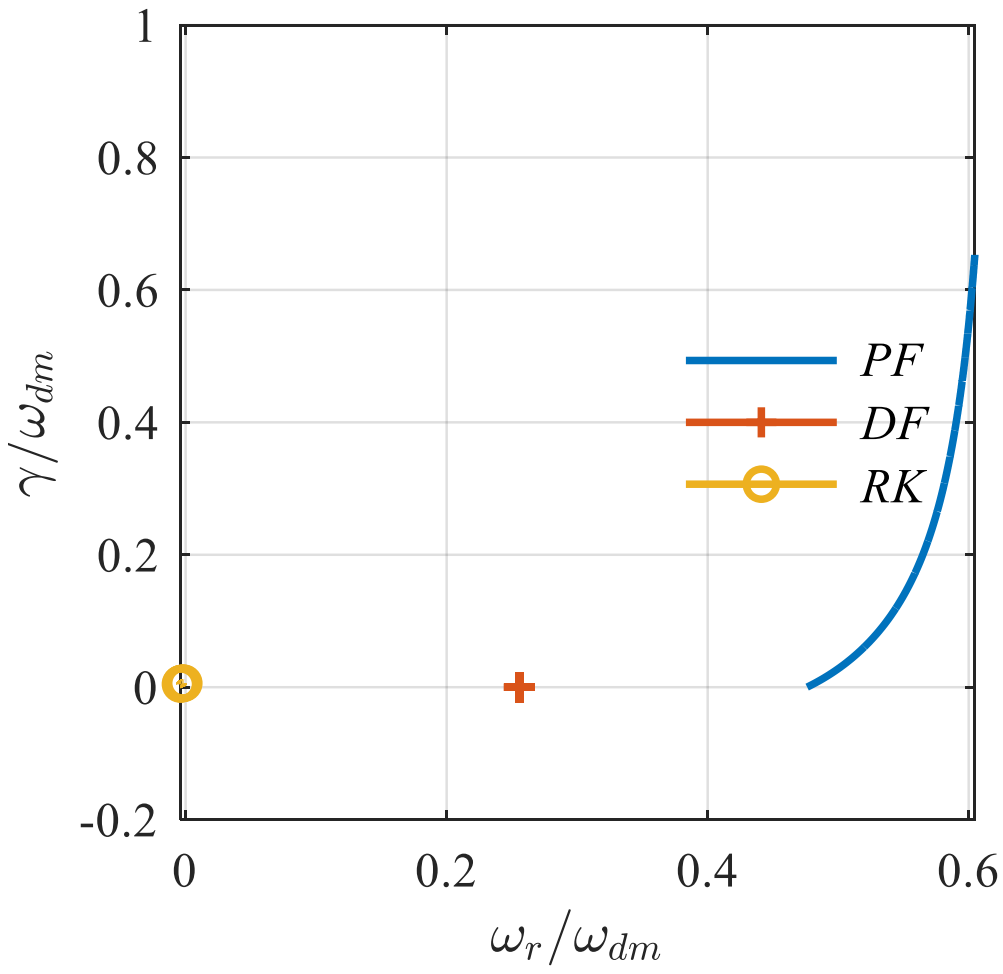

Figure 12. The three solutions of interest in the $(\omega_r, \gamma)$ plane (ranging $\beta_h$ from 0.00 to 0.01)

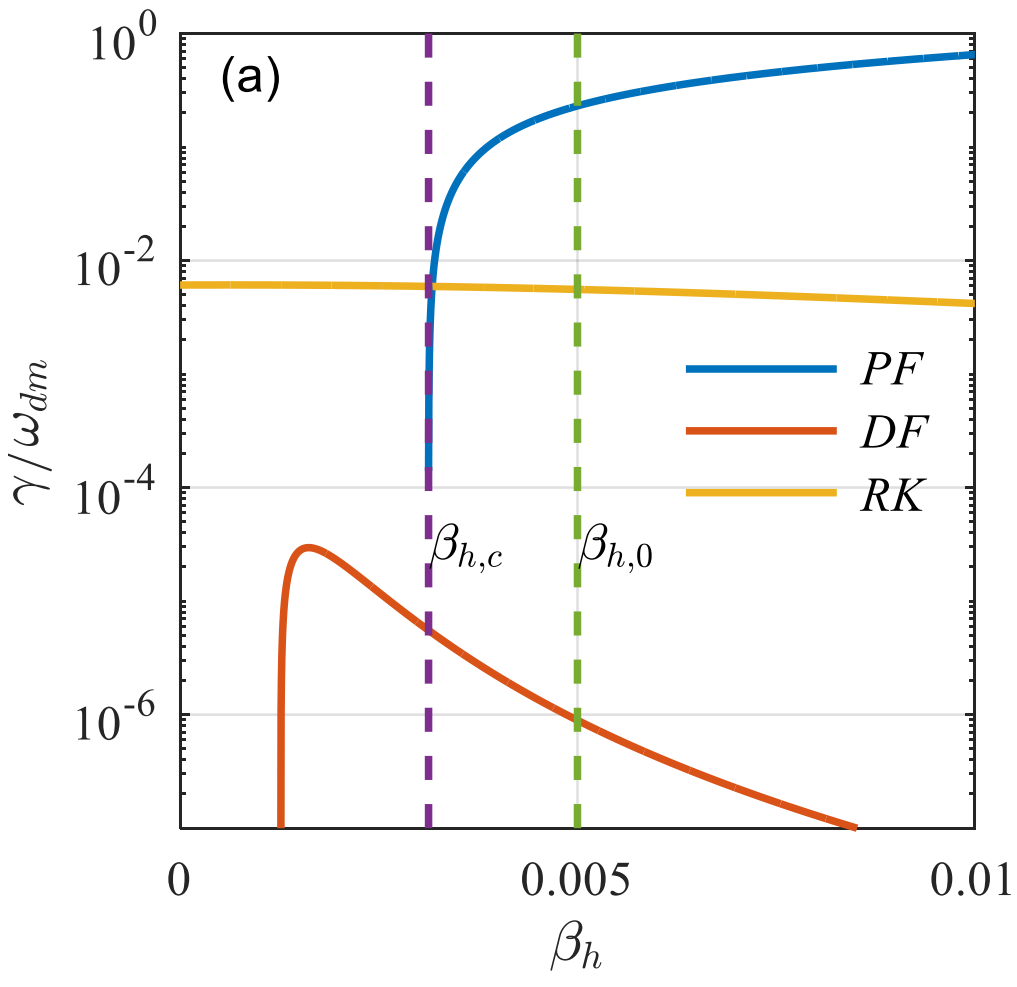

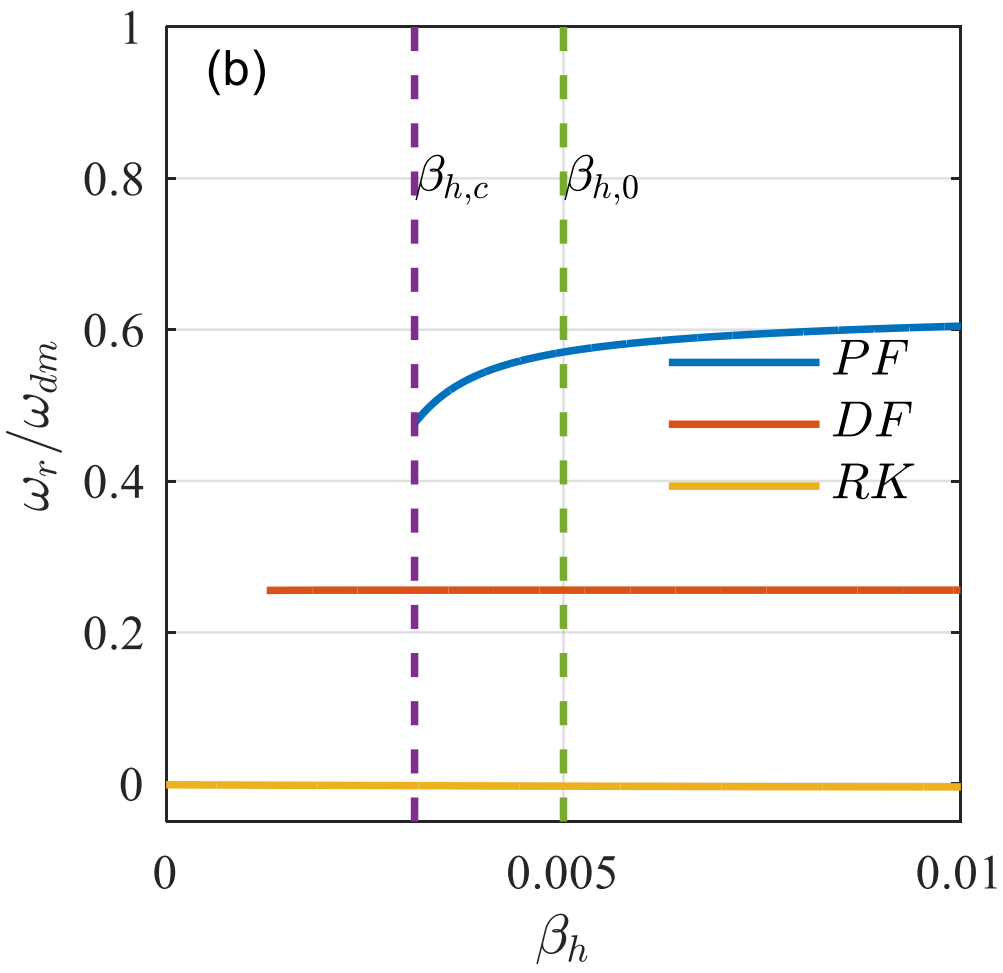

Figure 13. (a)The growth rate and (b) the real frequency of the modes versus $\beta_h$; $\beta_{h,c}$ denotes the threshold beta value of precession fishbone; $\beta_{h,0}$ means the beta value of NBI

## 5. Conclusion

| Range of $\omega_{*i}$ | $\omega_{*i} < \omega_{dm}/2$ | | | | $\omega_{dm}/2 < \omega_{*i} < \omega_{dm}$ | |
|---|---|---|---|---|---|---|
| Root | Diamagnetic Fishbone | | Precession Fishbone | | Diamagnetic Fishbone | |
| Real frequency | $\omega_r \approx \omega_{*i}$ | | $\omega_r \approx \omega_{dm}/2$ | | $\omega_r \approx \omega_{*i}$ | |
| Stability | $\lvert \Lambda^{3/2} \rvert \ll 1$ Weakly unstable | $\lvert \Lambda^{3/2} \rvert \gg 1$ Unstable | $\beta_h < \beta_{h,c}$ Stable | $\beta_h > \beta_{h,c}$ Unstable | $\lvert \Lambda^{3/2} \rvert \ll 1$ Weakly unstable | $\lvert \Lambda^{3/2} \rvert \gg 1$ Unstable |

Table 1. Features of Fishbone modes in different conditions

According to our calculation in Figures 10 and 11, assuming large pressure gradient and flat $q$-profile, the limit $|\Lambda^{3/2}| \gg 1$ for Diamagnetic Fishbone can be easily achieved. The condition could be roughly described as $(\omega_{A\theta}/S_M\omega_{*i})^{1/2} < Im(\delta\widehat{W}_k) + \delta\widehat{W}_c < \omega_{*i}/\omega_{A\theta}$ . Under the condition, mode with a real frequency $\omega_r \approx \omega_{*i}$ will be excited by the energetic trapped ions. At the same time, the root with $\omega_r \approx \omega_{dm}/2$ vanishes (Figure 7). In case of $\omega_{*i} < \omega_{dm}/2$, real frequency of the mode varies between $\omega_{*i}$ and $\omega_{dm}/2$ while in the case $\omega_{dm}/2 < \omega_{*i} < \omega_{dm}$, real frequency of the mode varies between $\omega_{*i}$ and $\omega_{dm}$, suggests that the resonant between diamagnetic oscillation and m=1 mode as well as resonant between trapped ions and m=1 mode is making contributions to the growth of the mode. Assuming $\omega_{*i} = \omega_{dm}/2$, the two roots nearly merge to one solution, both with $\omega_r = \omega_{dm}/2$ and a small growth rate $\gamma \approx O(10)/s$.

In conclusion, a detailed analysis is carried out for m/n=1/1 instabilities in one EAST shot via the combined fishbone dispersion relation. Three solutions are calculated for the specified discharge #48605. Precession fishbone is found unstable since the energetic ion beta exceeds the critical value, which is slightly affected by ion diamagnetic frequency. The Growth rate of the diamagnetic branch is much smaller in comparison with the precession branch. These results verify the theories of Chen and Coppi et al. For the third solution in the dispersion relation: resistive kink mode, the frequency and growth rate are also calculated. By comparison, it is shown that the precession fishbone has a much larger growth rate than the other two in a realistic beta range. However, the diamagnetic branch might be relevant for flat q profile or large diamagnetic frequency discharge.

# Appendix

## A1 The $\omega_{r,c}$ for Precession Fishbone in ideal limit

In the ideal limit, we have $S_M \gg 1$ as well as $\Lambda \gg 1$. Consider the Stirling Approximation in the limit $z \to \infty$ [14]

$$\Gamma(z) \approx \sqrt{2\pi(z-1)}\left(\frac{z-1}{e}\right)^{z-1}$$

$$ln[\Gamma(z)] \approx \frac{1}{2}ln2\pi + \frac{1}{2}lnz + zlnz - z$$

For $\alpha \in \mathbb{C}$, we have

$$ln\left[\frac{\Gamma(z+\alpha)}{\Gamma(z)}\right] \approx \frac{1}{2}ln\frac{z+\alpha}{z} + (z+\alpha)ln(z+\alpha) - zlnz - \alpha$$

Taking Taylor expansion in the limit $1/z \to 0$

$$ln\left[\frac{\Gamma(z+\alpha)}{\Gamma(z)}\right] \approx \alpha lnz + \frac{(\alpha+\alpha^2)}{2z} + O\left(\frac{1}{z}\right)^2 \approx \alpha lnz$$

$$\frac{\Gamma(z+\alpha)}{\Gamma(z)} \approx z^\alpha$$

Then we have

$$\frac{\Gamma[(\Lambda^{3/2}+5)/4]}{\Gamma[(\Lambda^{3/2}-1)/4]} \approx \frac{1}{8}\Lambda^{9/4}$$

The inertia $\delta\hat{I}$ becomes

$$\delta\hat{I} = -i\frac{[\omega(\omega-\omega_{*i})]^{\frac{1}{2}}}{\omega_{A\theta}}$$

Coupling equation (6) and (7), we have

$$\omega_{r,c}^2\left[ln\left(\frac{\omega_{dm}}{\omega_{r,c}}-1\right)\right]^2\left(1-\frac{\omega_{*i}}{\omega_{r,c}}\right) = \left(\delta\widehat{W}_c\pi\omega_{A\theta}\right)^2$$

Expansion near $\omega_{r,c} = \omega_{dm}/2$ gives $ln\left(\frac{\omega_{dm}}{\omega_{r,c}}-1\right) \approx 2\left(1-2\frac{\omega_{r,c}}{\omega_{dm}}\right)$, then we have

$$A^2\omega_{r,c}^2\left(1-A\omega_{r,c}\right)^2\left(1-\frac{B}{\omega_{r,c}}\right) = C$$

Where $A = 2/\omega_{dm}$, $B = \omega_{*i}$, $C = \left(\frac{\delta\widehat{W}_c\pi\omega_{A\theta}}{\omega_{dm}}\right)^2$. Considering $C \ll 1$, we find an asymptotic solution of the equation regarding the Precession Fishbone, which reads

$$\omega_{r,c} = \frac{1}{A}\left(1-\sqrt{\frac{C}{1-AB}}\right) + O(C)^2 \approx \frac{\omega_{dm}}{2}\left(1-\frac{\left|\delta\widehat{W}_c\pi\omega_{A\theta}/\omega_{dm}\right|}{\sqrt{1-2\omega_{*i}/\omega_{dm}}}\right)$$

Assuming $C = 0$， the solution reduces to $\omega_{r,c} = \frac{\omega_{dm}}{2}$.

## A2 The $\beta_{h,c}$ for Precession Fishbone in ideal limit

The dispersion relation then reads

$$[\omega(\omega-\omega_{*i})]^{1/2} = i\omega_{A\theta}\left[-\delta\widehat{W}_c - \beta_h\frac{\omega}{\varepsilon\omega_{dm}}ln\left(1-\frac{\omega_{dm}}{\omega}\right)\right]$$

Defining $\gamma_{mhd} = -\delta\widehat{W}_c$, equation (7) gives

$$e^{\frac{\gamma_{mhd}\varepsilon\omega_{dm}}{\beta_{h,c}\omega_r}} = \frac{\omega_{dm}}{\omega_r} - 1$$

For the realistic order that $\gamma_{mhd} \sim O(10^{-3})$, $\varepsilon \sim O(10^{-2})$, and $\beta_{h,c} \sim O(10^{-3})$

$$e^{\frac{\gamma_{mhd}\varepsilon\omega_{dm}}{\beta_{h,c}\omega_r}} = \frac{\omega_{dm}}{\omega_r} - 1 \approx 1$$

So, for ideal precession fishbone mode, we have $\omega_r \approx \omega_{dm}/2$.

Coupling equation (6), we have

$$\frac{\omega_{*i}}{1-\left(\frac{\pi\beta_{h,c}\omega_{A\theta}}{\varepsilon\omega_{dm}}\right)^2} = \frac{\omega_{dm}}{2}$$

It gives

$$\beta_{h,c} = \frac{\varepsilon\omega_{dm}}{\pi\omega_A}\left[1-\frac{2\omega_{*i}}{\omega_{dm}}\right]^{1/2}$$

### A3 The growth rate in ideal limit

Assuming $\omega_{*i} = 0$, the ideal dispersion relation becomes:

$$\omega = i\omega_{A\theta}\left[-\delta\widehat{W}_c - \beta_h \frac{\omega}{\varepsilon\omega_{dm}} ln\left(1-\frac{\omega_{dm}}{\omega}\right)\right]$$

In the marginal stability condition i.e. $\gamma = 0$, the imaginary part gives

$$\frac{\varepsilon\omega_{dm}}{\pi\omega_{A\theta}} = \beta_{h,c}$$

For the realistic order, we have $\left|\delta\widehat{W}_c\pi\omega_{A\theta}/\omega_{dm}\right| \ll 1$, the dispersion relation reduce to

$$\varepsilon\omega_{dm} = -i\omega_{A\theta}\beta_h\, ln\left(1-\frac{\omega_{dm}}{\omega}\right)$$

Coupling the two equations, we have

$$ln\left(1-\frac{\omega_{dm}}{\omega}\right) = i\pi\beta_{h,c}/\beta_h$$

Defining $\Omega = \Omega_R + i\Omega_I$, with $\Omega_R,\ \Omega_I$ denoting the real part and imaginary part of $\frac{\omega}{\omega_{dm}}$

$$ln\left(1-\frac{1}{\Omega_R + i\Omega_I}\right) = i\pi\beta_{h,c}/\beta_h$$

$$1-\frac{1}{\Omega_R + i\Omega_I} = e^{i\pi\beta_{h,c}/\beta_h}$$

Separating into real part and imaginary part

$$\frac{i\Omega_I}{\Omega_R^2 + \Omega_I^2} = isin\left(\pi\beta_{h,c}/\beta_h\right)$$

$$1-\frac{\Omega_R}{\Omega_R^2 + \Omega_I^2} = cos\left(\pi\beta_{h,c}/\beta_h\right)$$

The solution is

$$\Omega_I = \frac{1}{2}\frac{sin\left(\pi\beta_{h,c}/\beta_h\right)}{1-cos\left(\pi\beta_{h,c}/\beta_h\right)}$$

Its leading order is

$$\Omega_I \cong \frac{1}{2}\left[\frac{\pi\left(\beta_h - \beta_{h,c}\right)}{2\beta_{h,c}}\right]$$

Then we have the growth rate

$$\gamma_p \cong \frac{\pi^2\omega_{A\theta}}{4\varepsilon}\left(\beta_h - \beta_{h,c}\right)$$

## References

[1] K. McGuire, R. Goldston, M. Bell, M. Bitter, K. Bol, K. Brau, D. Buchenauer, T. Crowley, S. Davis, F. Dylla, H. Eubank, H. Fishman, R. Fonck, B. Grek, R. Grimm, R. Hawryluk, H. Hsuan, R. Hulse, R. Izzo, R. Kaita, S. Kaye, H. Kugel, D. Johnson, J. Manickam, D. Manos, D. Mansfield, E. Mazzucato, R. McCann, D. McCune, D. Monticello, R. Motley, D. Mueller, K. Oasa, M. Okabayashi, K. Owens, W. Park, M. Reusch, N. Sauthoff,

G. Schmidt, S. Sesnic, J. Strachan, C. Surko, R. Slusher, H. Takahashi, F. Tenney, P. Thomas, H. Towner, J. Valley, and R. White, *Study of High-Beta Magnetohydrodynamic Modes and Fast-Ion Losses in PDX*, Phys. Rev. Lett. **50**, 891 (1983).

[2] L. Chen, R. B. White, and M. N. Rosenbluth, *Excitation of Internal Kink Modes by Trapped Energetic Beam Ions*, Phys. Rev. Lett. **52**, 1122 (1984).

[3] H. Biglari and L. Chen, *Influence of Resistivity on Energetic Trapped Particle-induced Internal Kink Modes*, 6 (n.d.).

[4] B. Coppi and F. Porcelli, *Theoretical Model of Fishbone Oscillations in Magnetically Confined Plasmas*, Phys. Rev. Lett. **57**, 2272 (1986).

[5] B. Coppi, S. Migliuolo, and F. Porcelli, *Macroscopic Plasma Oscillation Bursts (Fishbones) Resulting from High-Energy Populations*, Phys. Fluids **31**, 1630 (1988).

[6] L. Xu, J. Zhang, K. Chen, L. Hu, E. Li, S. Lin, T. Shi, Y. Duan, and Y. Zhu, *Fishbone Activity in Experimental Advanced Superconducting Tokamak Neutral Beam Injection Plasma*, Phys. Plasmas **22**, 122510 (2015).

[7] G. Ara, B. Basu, and B. CoPPI, *Magnetic Reconnection and m = 1 Oscillations in Current Carrying Plasmas*, 34 (n.d.).

[8] F. Pegoraro and T. J. Schep, *Theory of Resistive Modes in the Ballooning Representation*, Plasma Phys. Control. Fusion **28**, 647 (1986).

[9] M. N. Bussac, R. Pellat, D. Edery, and J. L. Soule, *Internal Kink Modes in Toroidal Plasmas with Circular Cross Sections*, Phys. Rev. Lett. **35**, 1638 (1975).

[10] R. B. White, L. Chen, F. Romanelli, and R. Hay, *Trapped Particle Destabilization of the Internal Kink Mode*, 10 (n.d.).

[11] R. B. White, M. N. Bussac, and F. Romanelli, *High- β , Sawtooth-Free Tokamak Operation Using Energetic Trapped Particles*, Phys. Rev. Lett. **62**, 539 (1989).

[12] R. B. White, F. Romanelli, and M. N. Bussac, *Influence of an Energetic Ion Population on Tokamak Plasma Stability*, Phys. Fluids B Plasma Phys. **2**, 745 (1990).

[13] S. Bing-ren, J. W. VanDam, R. Carrera, and Y. Z. Zhang, *Resistive Effect on Ion Fishbone Mode in Tokamak Plasma*, Acta Phys. Sin. Overseas Ed. **2**, 260 (1993).

[14] M. Abramowitz and I. A. Stegun, editors , *Handbook of Mathematical Functions: With Formulas, Graphs, and Mathematical Tables*, 9. Dover print.; [Nachdr. der Ausg. von 1972] (Dover Publ, New York, NY, 2013).

[15] S. Migliuolo, *Theory of Ideal and Resistive M=1 Modes in Tokamaks*, Nucl. Fusion **33**, 1721 (1993).

[16] S. Migliuolo, F. Pegoraro, and F. Porcelli, *Stabilization of Collisional Drift-tearing Modes at the Breakdown of the Constant-Ψ Approximation*, Phys. Fluids B Plasma Phys. **3**, 1338 (1991).

[17] W. Shen, F. Wang, G. Y. Fu, L. Xu, G. Li, and C. Liu, *Hybrid Simulation of Fishbone Instabilities in the EAST Tokamak*, Nucl. Fusion **57**, 116035 (2017).